\documentclass[epjc3,smallextended,final,twocolumn]{svjour3} 

\usepackage{amsmath,amssymb,amsfonts}
\usepackage{bm,bbm}
\usepackage{graphicx}
\usepackage{color}
\usepackage{hyperref}

\newcommand{\be}{\begin{equation}}
\newcommand{\ee}{\end{equation}}
\newcommand{\ba}{\begin{eqnarray}}
\newcommand{\ea}{\end{eqnarray}}
\def\bs{\begin{subequations}}
\def\es{\end{subequations}}
\def\a{\alpha}
\def\b{\beta}
\def\de{\delta}

\def\k{\kappa}

\def\vr{\varrho}
\def\vp{\varphi}

\def\cT{\mathcal{T}}

\def\p{\partial}

\def\B{\Box}
\newcommand{\Eq}[1]{(\ref{#1})}
\def\com{\color{magenta}}
\def\cob{\color{blue}}

\newcommand{\oarX}[1]{\href{http://arxiv.org/abs/#1}{{\ttfamily\com #1}}}
\newcommand{\arX}[1]{\href{http://arxiv.org/abs/#1}{{\ttfamily\com arXiv:#1}}}
\newcommand{\doin}[6]{\href{http://dx.doi.org/#1}{\cob {#2 #3} {\bf #4}, #5 (#6)}}
\newcommand{\ndoin}[6]{\href{#1}{\cob {#2 #3} {\bf #4}, #5 (#6)}}
\newcommand{\tia}[1]{#1.}
\newcommand{\boxd}[1]{\boxed{#1}}

\def\rme{\text{e}}
\def\rmd{\text{d}}
\def\rmi{\text{i}}

\journalname{Eur. Phys. J. C}

\begin{document}

\title{Super-accelerating bouncing cosmology in asymptotically free non-local gravity}

\author{Gianluca Calcagni\thanksref{e1,addr1} \and Leonardo Modesto\thanksref{e2,addr2} \and Piero Nicolini\thanksref{e3,addr3}}

\thankstext{e1}{e-mail: calcagni@iem.cfmac.csic.es}
\thankstext{e2}{e-mail: lmodesto@fudan.edu.cn}
\thankstext{e3}{e-mail: nicolini@fias.uni-frankfurt.de}

\institute{Instituto de Estructura de la Materia, CSIC, Serrano 121, 28006 Madrid, Spain\label{addr1}
\and \noindent Department of Physics and Center for Field Theory and Particle Physics, Fudan University, 200433 Shanghai, China \label{addr2}
\and \noindent Frankfurt Institute for Advanced Studies (FIAS) and Institut f\"{u}r Theoretische Physik, Johann Wolfgang Goethe-Universit\"{a}t, 60438 Frankfurt am Main, Germany\label{addr3}}

\date{June 22, 2013}
\maketitle

\begin{abstract}
Recently, evidence has been collected that a class of gravitational theories with certain non-local operators is renormalizable. We consider one such model which, at the linear perturbative level, reproduces the effective non-local action for the light modes of bosonic closed string-field theory. Using the property of asymptotic freedom in the ultraviolet and fixing the classical behavior of the scale factor at late times, an algorithm is proposed to find general homogeneous cosmological solutions valid both at early and late times. Imposing a power-law classical limit, these solutions (including anisotropic ones) display a bounce, instead of a big-bang singularity, and super-accelerate near the bounce even in the absence of an inflaton or phantom field.
\end{abstract}

\keywords{Models of Quantum Gravity \and Cosmology of Theories Beyond the SM \and Spacetime Singularities}



\section{Introduction}

Asymptotic freedom is an attribute of field theories such that interactions are negligible in the ultraviolet (UV), where the theory possesses a trivial fixed point. This property has been used, directly or implicitly, to construct field theories of gravity where the Laplace--Beltrami operator $\B$ is replaced by one or more non-local operators $f(\B)$ in the kinetic terms of the action.  On a Minkowski background, if the Fourier transform $\tilde f(k^2)\to\infty$ for large momenta $k$, these terms dominate in the UV over the interactions, which can be ignored. Asymptotic freedom then, if realized, ensures that the correct UV behavior of the theory be encoded in the free propagator. A class of these models is of particular relevance inasmuch as the kinetic operator is of the exponential form $\B \rme^\B$ or $\rme^\B$ and is inspired, respectively, by open string-field theory (OSFT; see \cite{ohm01,Sen04,FK,Oka12} for reviews) and the $p$-adic string \cite{FO,FW,BFOW,VV,Vla07,BCK1}. In the simplest classical cosmological applications, gravity is local and the only non-local content is a scalar field \cite{are04,AJ,AKV1,cutac,AK,AV,BBC,kos07,AJV,AV2,lid07,BC,cuta2,Jou07,cuta4,Jo081,ArK,Jo082,BK2,NuM,KV1,cuta6,Ver09,Ver10,KV2}, sometimes identified with the tachyon of bosonic OSFT or of super-symmetric OSFT on an unstable brane.

Non-local gravity sectors and their cosmology have been proposed in \cite{Kra87,Tom97,ADDG,Bar1,Bar2,HaW,BMS,kho06,cuta8,BKM1,Bar3,Mod11,BGKM,Kos11,Mo12a,KV3,AMM,Mo12b,Mo12c,BKMV,BMMS,Kos13}, following various criteria including avoidance of ghosts, improved renormalizability, and the possibility to construct non-perturbative solutions. The gathered results (also in theories without non-local operators at the tree level \cite{HaM}) point towards a resolution of gravitational singularities thanks to asymptotic freedom. These approaches do not stem from closed string-field theory, i.e., the SFT sector containing the perturbative graviton mode \cite{SZ,KKS,KuS,KS3,Zwi92,SZ1,SZ2,SZ3,OZ,YZ1,YZ2,Mic06,Moe1,Moe2,Moe3}. Yet, many of them are inspired by it in the sense that the kinetic functions are exponential (or somewhat more general) operators as in effective closed SFT (see also \cite{NO,CENO,NOSZ,ZS,EPVZ} for other types of non-locality). In this context, one bypasses the technical difficulties in getting effective non-local actions directly from SFT \cite{KS3,Moe3} and concentrates on phenomenological but more manageable models.

Cosmological dynamical solutions were obtained either directly, by solving the non-local equations of motion, or indirectly, by solving an \emph{Ansatz} for the Ricci curvature (or via the diffusion equation \cite{cuta8}) respecting the equations of motion. Here we pursue a different but no less economic alternative, using asymptotic freedom as a key ingredient.

The strategy is the following. {\it (i)} First, in Sect.\ \ref{sec2} we define a model of non-local gravity inspired by closed SFT and falling into the class of actions considered in \cite{BKM1,Mod11}. {\it (ii)} In the UV, all interactions can be ignored and, thus, it is sufficient to find cosmological backgrounds compatible with the Green equation for the propagator. The scale factors $a(t)$ representing such backgrounds do not collapse into a big bang but, rather, display a bounce. At large scales, they reduce to known profiles of ordinary cosmology, the details of which depend on the choice of matter content. In particular, in Sect.\ \ref{sec3} we will find, with two different methods, backgrounds following a power law at late times. These profiles are approximate solutions of the full equations of motion valid both at very early times (when, roughly, the cosmological horizon scale is near the UV asymptotically free fixed point) and at late times. All such solutions have a bounce  and accelerate near it without invoking inflaton-like matter content.\footnote{See \cite{ABM,Rin09,BM2} for inflation-without-inflaton scenarios descending from other mechanisms.} {\it (iii)} Next, we write down an effective Friedmann equation of the form
\be\label{freq}
H^2= \frac{\k^2}{3}\rho_{\rm eff}:=\frac{\k^2}{3} \rho\left[1- \left(\frac{\rho}{\rho_*}\right)^\b\right],
\ee
where $H:=\dot a/a= \p_t a/a$ is the Hubble parameter, $\k^2=8\pi G$, $G$ is Newton's constant, $\rho$ is the energy density of the universe, $\rho_*$ is the critical energy density at which the bounce occurs, and $\b>0$ is a real parameter. The exponent $\b$ is determined by plugging the profile $a(t)$ found under the provision of asymptotic freedom into Eq.\ \Eq{freq} for a given energy density profile $\rho(a)$. Since this fitting is generally rather good, we can conclude that the class of asymptotic solutions found in step {\it (ii)} is reasonably valid also at intermediate times, and that the bouncing-accelerating scenario of the theory is well described by the effective Friedmann equation \Eq{freq}.


\section{The model}\label{sec2}


\subsection{Effective action from closed string-field theory}

We start by recalling the derivation of the effective SFT Lagrangian for the closed string through a mass-level truncation scheme. In the non-polynomial bosonic closed SFT, the action has the following compact form:
\be
S_{\rm SFT}  =  \int   \! \frac{1}{\alpha^\prime} \Phi * Q b_0^- \Phi + g \! \sum_{N = 3}^{+ \infty} 
\frac{ (g \, \alpha^{\prime} )^{N-3}  }{2^{N-3} N ! } 
\Phi * [ \Phi^{N-1}]\,,
\ee
where $\Phi$ is the closed-string field, $Q$ is the BRST operator, $b_0^- = (b_0 - \bar{b}_0)/2$
is a combination of antighost zero modes, $\a'$ is the Regge slope and $g$ is the string-field coupling constant.
The integral $\int \Phi_1 * \Phi_2$ represents the string-field scalar product, while the symbol $[\Phi^{N-1}]=[\Phi_{1}\dots\Phi_{N-1}]$ denotes the string field obtained combining $N-1$ fields $\Phi_{1},\dots,\Phi_{N-1}$ using the $N$-string vertex function.

The idea that light states dominate physical processes justifies the following truncation of the string field in terms of oscillators and particle fields:
\ba
\Phi &=& {c}_0^- \! \left[ \phi + A_{\mu \nu} \, \alpha_{-1}^{\mu}  \bar{\alpha}_{-1}^{\nu} +
\frac{ (\alpha_+ + \alpha_-) b_{-1} \bar{c}_{-1} }{\sqrt{2}}\right.\nonumber\\
&&\qquad+\frac{ (\alpha_+ - \alpha_-) {c}_{-1}\bar{b}_{-1}  }{\sqrt{2}}\nonumber\\
&&\qquad
 \left.+  \vphantom{\frac{ \bar{c}_{-1} }{\sqrt{2}}} \rmi c_0^+ \, ( j_{1 \mu} \alpha^{\mu}_{-1} \bar{b}_{-1} + 
  j_{2 \mu} {b}_{-1} \bar{\alpha}^{\mu}_{-1} ) \right]  \! | 0 \rangle, \label{tru}
\ea
where $A_{(\mu \nu)} := (A_{\mu \nu} + A_{\nu \mu})/2$ is the graviton field, $A_{[\mu \nu]} := (A_{\mu \nu} - A_{\nu \mu})/2$ is an antisymmetric rank-2 tensor field, $\alpha_{\pm}$, $j_{1 \mu}$ and $j_{2 \mu}$ are auxiliary fields, and $b$ and $c$ are first-quantized ghost oscillators. Greek indices run over spacetime directions and are lowered via the Minkowski metric $\eta_{\mu\nu}={\rm diag}(+,-,\dots,-)$. The state $|0 \rangle = \bar{c}_1 | \bar{\Omega} \rangle \otimes {c}_1 | {\Omega} \rangle $ is the first-quantized string vacuum, with $|\bar{\Omega} \rangle$ and $| {\Omega} \rangle $ the left and right $SL(2,\mathbb{R})$-invariant vacua. Further details can be found, e.g., in \cite{ohm01}.

The truncation \Eq{tru} allows one to derive the cubic effective Lagrangian $\mathcal{L} = \mathcal{L}_{\rm free} + \mathcal{L}_{\rm int}$. Working in the Siegel--Feynman gauge $b_0^+ \Phi = 0$ (which sets $j_{1 \mu} = j_{2 \mu} =0$), the kinetic and mass terms read
\ba
\mathcal{L}_{\rm free} &=& \frac{1}{2} \partial_{\lambda} A_{\mu \nu} \partial^{\lambda} A^{\mu \nu} 
+ \frac{1}{2} \partial_{\lambda} \phi \partial^{\lambda} \phi + \frac{2}{\alpha^\prime} \phi^2\nonumber\\
&&+ \frac{1}{2} \partial_{\lambda} \alpha_+ \partial^{\lambda} \alpha_+ -\frac{1}{2} \partial_{\lambda} \alpha_- \partial^{\lambda} \alpha_-\,,
\ea
while $\mathcal{L}_{\rm int}=\mathcal{L}_{\rm int}(\p,\tilde\phi,\tilde A,\tilde\a_\pm)$ has a number of interaction terms with derivatives.
For a given field $\vp(x)$ in the kinetic term, we have the corresponding ``dressed'' field 
\be
\tilde{\vp}(x) = \rme^{-\Box/(2\Lambda^2)} \, \vp(x)\,, \qquad
\Box= \partial_{\mu} \partial^{\mu}
\ee
in the interaction part, where $1/\Lambda^2 =\alpha'\ln (3 \sqrt{3}/4)\approx 0.2616\,\alpha'$. Operators of this form are a fully non-per-turbative effect of the string interactions. The conformal field theory describing such interactions is based upon special Fock states which obey a universal diffusion equation; such a structure is inherited by the effective spacetime action, as a residual manifestation of OSFT gauge invariance \cite{cuta7}. Non-local operators are, in general, associated with extended objects (rather than pointwise particles). Specifically, the exponential operators of the type considered in this paper are an imprint of the finite size of the string \cite{CM1}.

The above Lagrangian ${\cal L}$ can be recast in an equivalent form by shifting the smearing functions from the interaction term to the kinetic and mass ones by the field redefinition $\vp(x) \rightarrow \rme^{\Box/(2\Lambda^2)} \, \vp(x)$. Ignoring the auxiliary fields, the free part of the Lagrangian reads 
\be
\mathcal{L}_{\rm free} = - \frac{1}{2}  A_{\mu \nu} \, \Box \, 
\rme^{\Box/\Lambda^2}  \, A^{\mu \nu} 
- \frac{1}{2} \phi \left( \Box  - \frac{2}{\alpha^{\prime} } \right) \rme^{\Box/\Lambda^2} \phi\,,
\label{kinnico} 
\ee
which leads to the following stringy modifications of the Laplace--Beltrami operator in the zero-level truncation scheme:
\be\label{stringprop}
\Box \,\, \rightarrow \,\,\Box\, \rme^{\Box/\Lambda^2}\,. 
\ee
The associated propagator generally leads to a ghost-free spectrum, the intuitive reason being that entire functions $f(\B)$ do not introduce extra poles \cite{cutac,cuta4,BMS,BK1}. Quantum field theories with exponential propagators have been argued long since to be super-renor-malizable \cite{Efi77}. Operators of the form \Eq{stringprop} also appear in the context of non-commutative geometries, where a minimal length is effectively induced \cite{SmS,SSN,NiR,KoN}.


\subsection{Non-local gravity model}

From the tree-level effective action for the graviton and matter, we can argue about the form of its non-linear covariant extension. On the ground of a recently introduced candidate model of super-renormalizable gravity \cite{BKM1,Mod11,BGKM,Mo12a,Mo12b,Mo12c} based on earlier results \cite{Kra87,Tom97} (see also \cite{ALS} for considerations in a local higher-derivative theory), we propose the action\footnote{This action differs from the theory of \cite{cuta8}, which is of scalar-tensor type. There, the curvature invariants appear only in the exponential non-local operator in order to allow for non-perturbative solutions via a method based on the diffusion equation.}
\bs\label{Seff}
\be
S = \frac{1}{2\k^2_D} \int \rmd^Dx\, \sqrt{|g|}\, \left[R- G_{\mu\nu}\, \gamma(\Box)\, R^{\mu\nu}\right]+S_{\rm matter}\,,\label{Sg}
\ee
\ba
S_{\rm matter} &=&  \int \rmd^Dx\, \sqrt{|g|}\, \left[\frac{1}{2}  \nabla_\mu \phi \, V^{-1}(\Box) \, \nabla^\mu \phi\right.\nonumber\\
&&\qquad\left.+ \frac{1}{2 n !} \rme^{c \phi} \, F_{[n]} \, V^{-1}(\Box)\, F_{[n]}  \right],
\ea\es
where $G_{\mu\nu}$ and $R_{\mu\nu}$ are, respectively, the Einstein tensor and the Ricci tensor associated with the $D$-dimen-sional target spacetime metric $g_{\mu\nu}$, and, adopting the terminology used in string theory, $\phi$ is the dilaton field coming from the trace of the field $A_{\mu\nu}$, $n = p+2$, and $F_{[n]}$ are the $p$-field strengths corresponding to the gauge potentials. The value of the parameter $c$ controls the interaction of the scalar field $\phi$ with the field strength $F_{[n]}$. The key ingredients of the above action are the operators
\be
\gamma(\Box) :=  \frac{V^{-1}(\Box) -1}{\Box}\,,\qquad V(\B) := \rme^{- \Box/\Lambda^2},\label{VSFT}
\ee
where $\Lambda$ (proportional to ${\a'}^{-1/2}$ in SFT) is the invariant energy scale above which quantum-gravity effects become non-negligible. This form of the kinetic terms correctly reproduces the non-local operator \Eq{stringprop} (and its inverse, the propagator $V(\B)/\B$) when linearizing the fields \cite{Kra87,Tom97}. It also leads to the improved renormalizability of the model. However, in a spacetime of even dimension the effective action is not generally finite but only super-renormalizable because one-loop diagrams are still superficially divergent \cite{Kra87,Tom97,Mod11,BGKM,Mo12a,Mo12b,Mo12c,ALS,MMN}. Things go differently in a spacetime of odd dimension, because at the one-loop level there are no local operators which can serve as counter-terms for pure gravity and the theory results to be finite \cite{FQG,MTheory}. When, in the case of super-symmetry, matter is added to fill up the super-gravity multiplet, the theory remains finite \cite{duff}.

From the point of view of string-field theory, the action \Eq{Seff} only contains massless fields, but in general the whole tower of string massive particle modes should be taken into account. However, for the cosmology-related purposes of the present paper the massless sector is more than enough, since gravity is included in it and, on the other hand, cosmological matter can be modeled by a perfect fluid as a first approximation, even if at the fundamental level it is constituted by fields with non-local dynamics. In this regard, notice that the form of the kinetic operators in \Eq{Seff} is expected to hold also for massive modes \cite{MTheory} (consult that reference for a longer discussion). One simple example of this is given by the string tachyon, which is a massive field although with negative squared mass; as one can see from Eq.\ \Eq{kinnico}, its kinetic operator is exactly of the same form as for the massless fields. Another motivation to drop massive states from the discussion is that a perturbative truncation of the non-local operators would correspond to a small Regge slope. By keeping these operators intact we imply no special requirement on the size of squared momenta $k^2$ with respect to $\a'$. In particular, one is entitled to explore configurations such that $\a'k^2\gg 1$. This is the region in parameter space where all the masses of the tower, which are multiples of the fundamental mass $1/\sqrt{\a'}$, can be ignored with respect to kinetic terms $\propto k^2\gamma(-k^2)$. Therefore, a truncation of the massive tower is in principle compatible with keeping the non-local operators untruncated.

In this paper, we study asymptotic profiles which are approximate cosmological solutions of the gravitational system \Eq{Sg}. The classical action \Eq{Sg} is a ``non-polynomial'' or ``semi-polynomial'' extension of quadratic Stelle theory \cite{ALS,Ste77}. All the non-polynomiality is incorporated in the form factor $\gamma(\Box)$. The entire function $V$ has no poles in the whole complex plane, which preserves unitarity, and it has at least logarithmic behavior in the UV regime to give super-renormalizability at the quantum level.

Here we only consider corrections to the classical solutions coming from the bare two-point function of the graviton field. The reason is that this class of theories is asymptotically free and the leading asymptotic behavior of the dressed propagator is dominated by its bare part. In fact, according to power counting arguments \cite{Tom97,Mod11}, the self-energy insertions, which are constant or at most logarithmic, do not contribute to it. 


\section{Cosmology}\label{sec3}


\subsection{General solution in asymptotically free gravity}\label{geso}

To find general homogeneous and isotropic cosmological solutions in the UV, we adapt a procedure used in ordinary perturbative quantum gravity \cite{Duf74,Bro10,Bro11}. Later on, we shall comment on the main differences of our setting with respect to earlier applications. The same method can be applied also in the case of gravitational collapse \cite{BMM}. We fix the number of dimensions to $D=4$.  Consider a homogeneous and isotropic Friedmann--Robertson--Walker (FRW) metric $g_{\mu \nu}$ with zero curvature; the anisotropic case is straightforward and will be discussed later. We split the metric into a flat Minkowski background plus a homogeneous tensor $h_{\mu \nu}$,
\bs\label{ansa}\ba
g_{\mu \nu} &=& \eta_{\mu \nu} + \kappa \, h_{\mu \nu} \,, \\ 
\rmd s^2 &=& g_{\mu \nu} \rmd x^{\mu} \rmd x^{\nu} = \rmd t^2 - a(t)^2 \,\delta_{i j}\rmd x^i \rmd x^j  \,,
\ea\es
where $i=1,2,3$. By definition, $h$ is small around a certain time $t_{\rm i}$, where $g_{\mu \nu}(t_{\rm i}) = \eta_{\mu \nu}$, but only then. In fact, the \emph{Ansatz} \Eq{ansa} is not to be understood in the sense of cosmological perturbation theory, where $\kappa h\ll 1$ is a perturbation small everywhere and at any time. Equation \Eq{ansa} is simply a splitting of the FRW metric, not the perturbation of Minkowski background with a generic fluctuation. Therefore, we stress that the present method has nothing to do with cosmological perturbation theory.\footnote{Throughout the paper, and unless stated otherwise, we will use the term ``perturbation theory'' exclusively in the sense of field theory, not of cosmology.}

Thus, the scale factor $a$ and the fluctuation $h_{\mu\nu}$ are
\bs\label{ah}\ba
&& a^2(t) = 1 - \kappa h(t) \,, \qquad h(t=t_{\rm i}) = 0 \,,\nonumber\\ 
&&\quad  g_{\mu \nu}(t=t_{\rm i}) = \eta_{\mu\nu} \,,\\
&& h_{\mu \nu}(t) = h(t) \, {\rm diag}(0, \delta_{i j}) =: h(t) \, \mathcal{I}_{\mu \nu} \,.
\ea\es
The tensor $h_{\mu \nu}$ can be rewritten in harmonic gauge by the transformation
\ba
&& h_{\mu \nu}(x) \rightarrow h^{\prime}_{\mu \nu}(x) := h_{\mu \nu}(x)+ \partial_{\mu} \xi_{\nu} + \partial_{\nu} \xi_{\mu}
\,, \nonumber\\
&&\xi_{\mu}(t) = - \frac{3 \kappa}{2} \, {\rm diag}\left[\int_{t_{\rm i}}^t dt'\,h(t'),0,0,0  \right]\!. \nonumber
\ea
The fluctuation now reads 
\be
h^{\prime}_{\mu \nu}(t) = 
h(t) \, {\rm diag} ( - 3, \delta_{i j} ) \,, \qquad h^{\prime \, \mu}_{\mu}(t) = - 6 h(t) \,.
\ee
The standard gravitational field $\bar{h}_{\mu \nu}$ is then 
\ba
\bar{h}_{\mu \nu} &:=& {h}^{\prime}_{\mu \nu} - \frac{1}{2} \eta_{\mu \nu} \, h^{\prime \, \lambda}_{\lambda}
= h(t) \, {\rm diag}(0, -2 \delta_{ij})\nonumber\\
&=& - 2h(t) \mathcal{I}_{\mu \nu}\,,\nonumber\\
\partial^{\mu} \bar{h}_{\mu \nu} &=& 0\,.
\ea
The Fourier transform of the above field is given by
\be
\tilde{\bar{h}}_{\mu \nu}(E, \vec{k}) = -2 \tilde{h}(E) (2 \pi)^3 \delta(\vec{k}) \, \mathcal{I}_{\mu \nu}\,. 
\label{FTh}
\ee

At this point, we exploit asymptotic freedom to avoid solving the full equations of motions of \Eq{Seff} on a flat FRW background, and to recognize the profile \Eq{ah} as an actual asymptotic {solution} of our model. We express the classical propagator for the excitation $\bar h_{\mu\nu}$ via a dimensionless source $\vr$, representing the 00 component of an effective ``energy-momentum'' tensor $\cT^{\mu\nu}$. Denote its Fourier transform with a tilde. The gauge-independent part of the graviton propagator \cite{Mod11,AAM} for the theory (\ref{Sg}) is then
\ba 
&& \mathcal{O}^{-1}(k) = \frac{V(k^2) }{k^2}   \left( P^{(2)} - \frac{P^{(0)}   }{2}  \right) 
\,\,\, \,\, 
\Longrightarrow \nonumber\\
&&
\bar{h}_{\mu \nu}(x) = \kappa \! \int \! \frac{d^4 k}{(2 \pi)^4}
\mathcal{O}^{-1}_{\mu \nu, \rho \sigma}(k) \tilde{\cT}^{\rho \sigma}(k) \, \rme^{-\rmi k\cdot x} \,,
\label{propgauge2} 
\ea
where $P^{(0)}$ and $P^{(2)}$ are Van Nieuwenhuizen projectors in four dimensions \cite{VN}. The standard ``classical'' case is obtained for $V(k^2)\to 1$, but for the theory (\ref{Sg}) we find, using the graviton propagator after a Wick rotation,
\ba
h(t) &=& -\frac{\kappa}{2} \! \int \frac{\rmd^4 k}{(2 \pi)^4} \frac{1}{k^2 V^{-1}(k^2)} \, \tilde{\vr}(E, \vec{k}) \,
\rme^{-\rmi k\cdot x}\nonumber\\
&=& \int \frac{\rmd E}{2 \pi}\, \tilde h(E)\,  \rme^{\rmi E t}\,.
\label{htg}
\ea
Assuming that the kernel $\tilde\vr$ does not depend on the cut-off $\Lambda$, the quantum-corrected profile $a(t)$ can be found from its ``classical'' limit $a_{\rm cl}(t)$. The procedure is the following: {\it (i)} fix an $a_{\rm cl}(t)$ and, via Eq.\ \Eq{ah}, a profile $h_{\rm cl}(t)$; {\it (ii)} set temporarily $V=1$ in Eq.\ \Eq{htg} and obtain $\tilde h(E)$ and, from that, the cut-off-independent distribution $\tilde{\vr}(E, \vec{k})$; 
 {\it (iii)} plug $\tilde{\vr}$ back into Eq.\ \Eq{htg} and perform the four-momentum integral with the full $V(k^2)\neq 1$, to obtain $h(t)$; {\it (iv)} use \Eq{ah} to get $a(t)$. The profile $a_{\rm cl}(t)$ is thus recovered perturbatively at times $t\gg t_{\rm i}$.

This procedure is similar to the one employed in \cite{Duf74,Bro10,Bro11}, but with an important difference. There, in order to go beyond the classical theory, one introduces one-loop quantum corrections to the graviton propagator. In our case, however, we already have modifications at the classical level and, therefore, we use only the bare propagator. This can be justified by noting, as mentioned above, that one-loop corrections to the propagator in this class of non-local theories are UV sub-dominant with respect to the tree-level contribution. Thus, a general conclusion is that any asymptotically free theory of gravity with a two-point function of the form (\ref{propgauge2}), with sufficiently strong damping factor $V$, will admit an asymptotic UV solution of the form \Eq{ah} solely found at the tree level in perturbation theory.


\subsection{General solution with power-law regime}

We start by considering a profile compatible, at late times, with a power law. We recall that we are not in vacuum and the matter source shapes the expansion of the universe. The classical profile is very simple, namely,
\be
a_{\rm cl}(t) = \left|\frac{t}{t_{\rm i}} \right|^p \,,\qquad
h_{\rm cl}(t) = \frac{1}{\kappa} \left[1 - \left|\frac{t}{t_{\rm i}} \right|^{2p} \right] \,,
\label{ah2}
\ee
where $t_{\rm i}$ is the pivot time around which one centers the splitting of the metric, $t=0$ is the big-bang singularity time, and $p>0$ is a constant. Equation \Eq{ah2} is solution to the ordinary Einstein equations, recovered at low curvature also in our model when higher-order Riemann terms are negligible. $a_{\rm cl}(t)$ is an even function of time due to time-reversal symmetry in the standard classical Friedmann equations and, in turn, it will determine an even quantum-corrected profile $a(t)=a(|t|)$. This will also prevent a possible issue with the procedure {\it (i)}--{\it (iv)} detailed above. The propagator is integrated after Wick rotation, when the form factor $V$ makes it convergent. For the purpose of field-theory calculations, going to imaginary time poses no particular problem, provided the Osterwalder--Schrader conditions are satisfied \cite[Section 5]{Tom97}. However, the solution $a(t)$ is found with the Wick-rotated propagator, and its analytic continuation back in Lorentzian time $t\to -\rmi t$ may no longer be sensible for the Lorentzian system. In the present case, however, the solution depends on $t^2=|t|^2$ ($t$ is a \emph{real} parameter) and it correctly reaches the classical power-law solution of the Lorentzian theory at late times. Therefore, the profile $a(t)$ is unaffected by the analytic continuation. In Sect.\ \ref{altder}, we will recover the same solution with an independent method. 

Given the condition \Eq{ah2}, we can calculate the Fourier transform $\tilde{h}(E)$ defined in Eq.\ (\ref{FTh}),
\be
\tilde{h}(E) = \frac{1}{\k E^2 V^{-1}(E^2)}\left[\pi E^2\delta (E)+ \frac{\sin(\pi p)\Gamma(2 p+1)}{t_{\rm i}^{2 p}|E|^{2p-1}}\right],
\ee
and, from Eq.\ \Eq{htg}, the source $\tilde\vr$ in momentum space:
\ba
\tilde{\vr}(E, \vec{k}) &=& \frac{4}{\k^2}\left[\pi E^2\delta (E)+ \frac{\sin(\pi p)\Gamma(2 p+1)}{t_{\rm i}^{2 p}|E|^{2p-1}}\right]\nonumber\\
&&\times (2 \pi)^3 \delta(\vec{k})\,.
\ea
For the form factor in Eq.\ (\ref{VSFT}), the fluctuation resulting from the integral (\ref{htg}) is
\be
\kappa \, h(t) = 1 - \left(\frac{2}{\Lambda t_{\rm i}}\right)^{2p}\frac{\Gamma\left(\frac12+p\right)}{\sqrt{\pi}} \, _1F_1\left(-p;\frac{1}{2};-\frac{1}{4} t^2 \Lambda ^2\right) \,,
\label{hsft}
\ee
where $ _1F_1$ is the confluent hypergeometric function of the first kind (Kummer's function):
\be\label{kumm}
_1F_1(b;\,c;\,z):=\frac{\Gamma(c)}{\Gamma(b)}\,\sum_{l=0}^{+\infty} \frac{\Gamma(b+l)}{\Gamma(c+l)}\, \frac{z^l}{l!}\,.
\ee
The transcendental equation \Eq{hsft} determines the value of $\Lambda t_{\rm i}$ such that $h(t_{\rm i}) = 0$. This only affects unimportant time rescalings and the normalization of the scale factor
\be\label{solu}
\boxd{a(t) =\left(\frac{2}{\Lambda t_{\rm i}}\right)^{p}\sqrt{\frac{\Gamma\left(\frac12+p\right)}{\sqrt{\pi}} \, _1F_1\left(-p;\frac{1}{2};-\frac{1}{4} t^2 \Lambda ^2\right)}\,.}
\ee
The early- and late-time behavior can easily be found from the asymptotics of Kummer's function (e.g., \cite{cuta2}):
\ba
_1F_1(b;\,c;\,-z^2) &\ \stackrel{\text{\tiny $z\to\pm \infty$}}{\sim}\ &\frac{\Gamma(c)}{\Gamma(c-b)} (z^2)^{-b}\,,\nonumber\\
_1F_1(b;\,c;\,-z^2) &\ \stackrel{\text{\tiny $z\to 0$}}{\sim}\ & 1\,.\label{largez}
\ea
We obtain
\ba
a(t) &\ \stackrel{\text{\tiny $t\to\pm\infty$}}{\sim}\ & \left|\frac{t}{t_{\rm i}}\right|^p, \nonumber\\
a(t) &\ \stackrel{\text{\tiny $t\to 0$}}{\sim}\ & a_*:=\sqrt{\frac{\Gamma\left(\frac12+p\right)}{\sqrt{\pi}}}\left(\frac{2}{\Lambda t_{\rm i}}\right)^p.\label{boun}
\ea
The transition between these two regimes is set by the critical time $1/\Lambda$. The general picture is that, at late times, the universe expands as a power law determined by its matter content. Common cases are radiation ($p=1/2$), dust matter ($p=2/3$), and acceleration-inducing components such as inflaton and quintessence ($p>1$). In all these scenarios, there is a finite bounce at $t=0$, whose value increases with $p$ if $\Lambda t_{\rm i}\sim 1$.  The most direct agent responsible for the bounce is asymptotic freedom: when the energy scale $\Lambda$ is sent to infinity, $a_*\to 0$.

\begin{figure}
\hspace{-0.4cm}
\centering
\includegraphics[width=7.0cm]{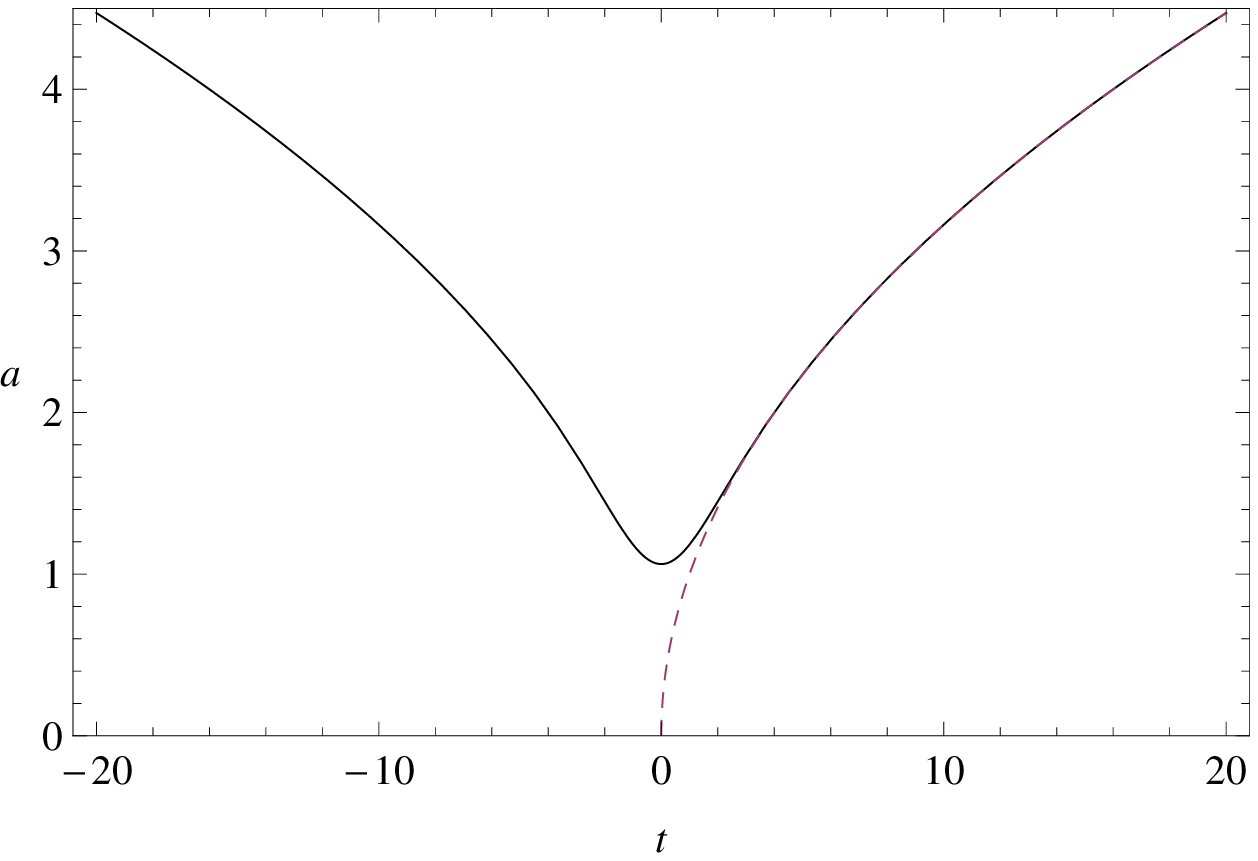}
\hspace{0.1cm}
\includegraphics[width=7.21cm]{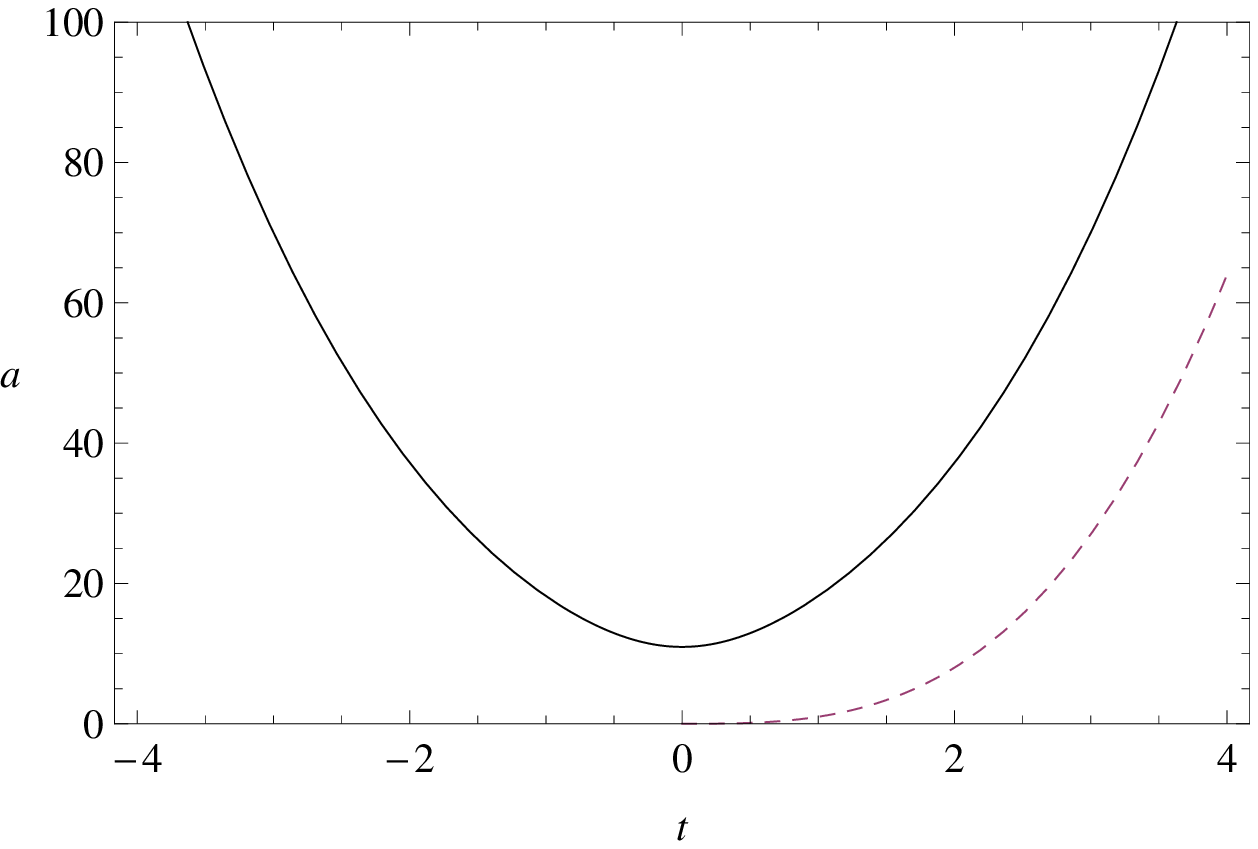}
\caption{\label{fig1} Typical bouncing profile of the scale factor for $p<1$ (upper plot, $p=1/2$, radiation) and $p>1$ (lower plot, $p=3$, accelerating universe), for $\Lambda=1=t_{\rm i}$. The convexity of the curve changes sign at $p=1$. Dashed curves show the corresponding power-law profiles $a\sim t^p$ with big-bang singularity.}
\end{figure}

\begin{figure}
\hspace{-0.4cm}
\centering
\includegraphics[width=7.3cm]{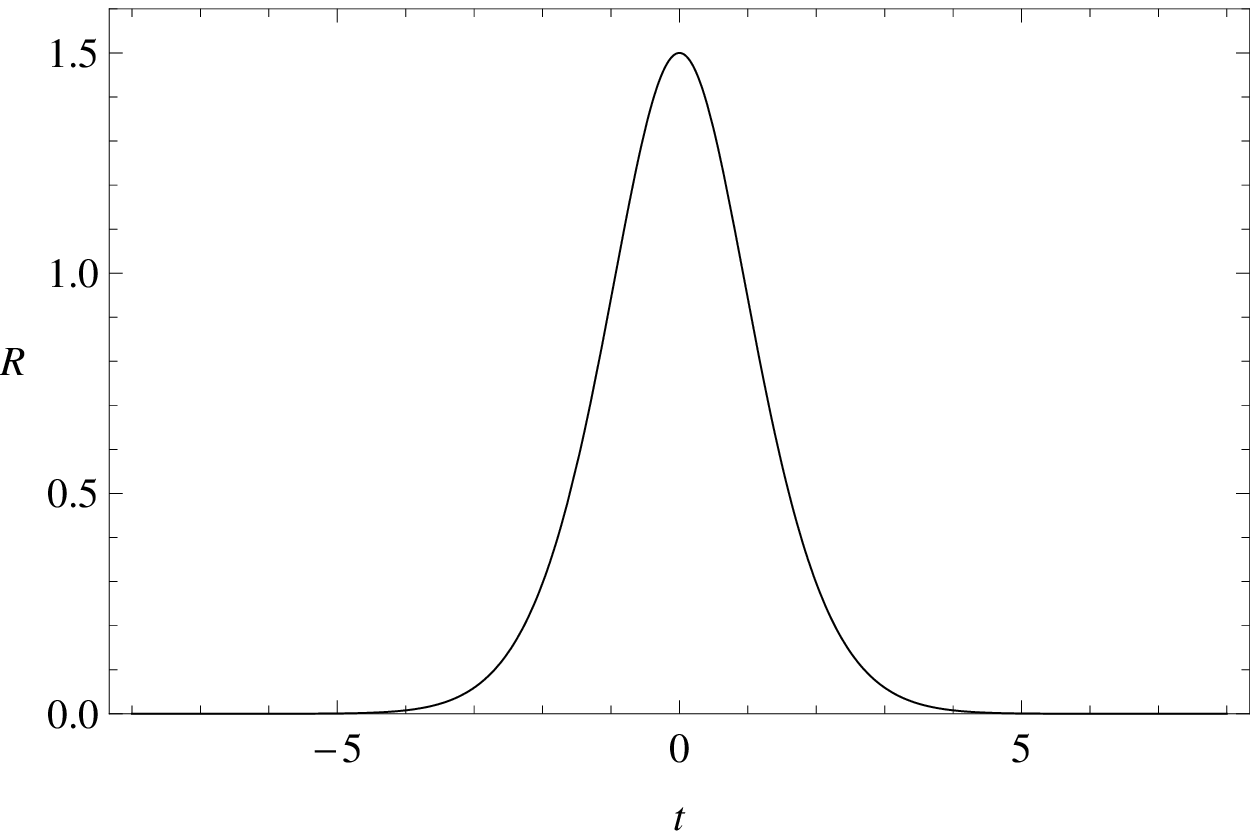}
\hspace{0.1cm}
\includegraphics[width=7.25cm]{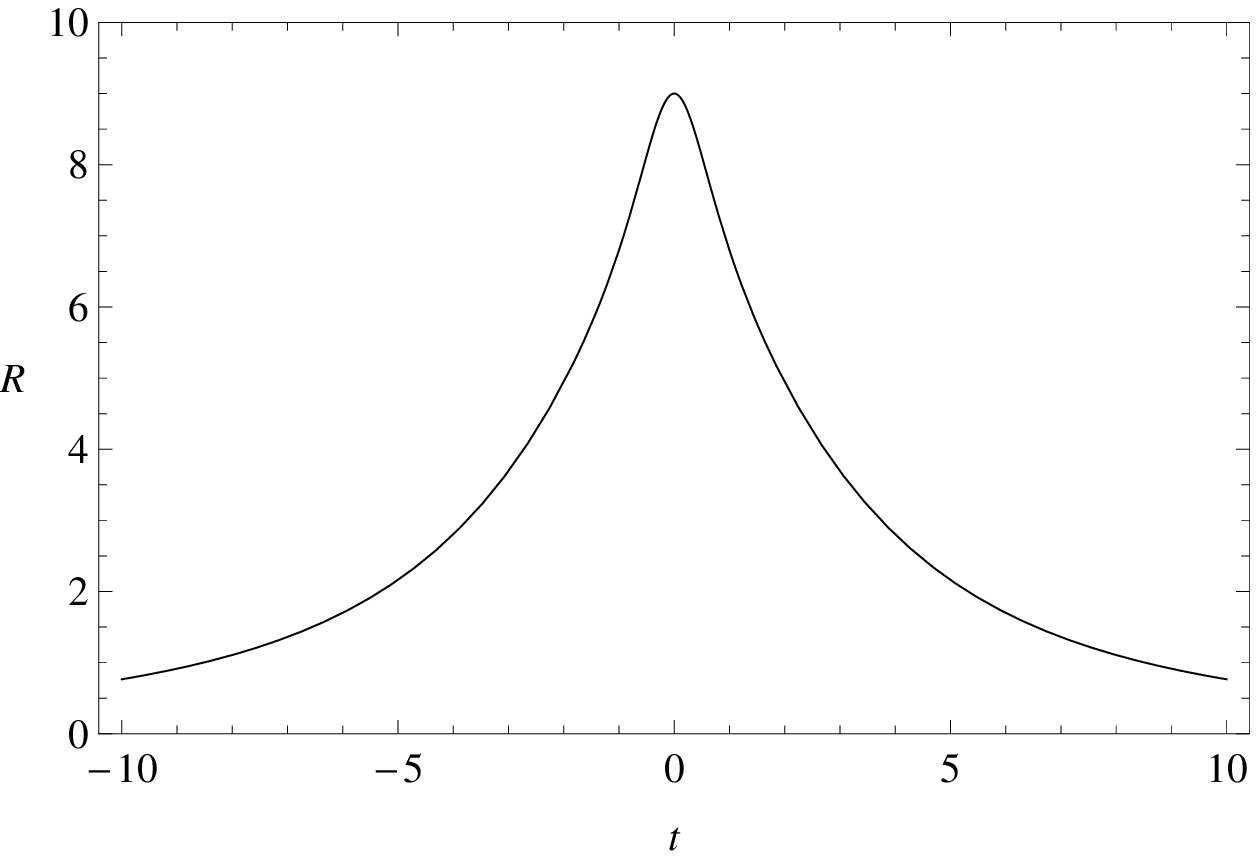}
\caption{\label{fig2} Ricci scalar $R$ for $p=1/2$ (top) and $p=3$ (bottom). The curvature increases towards the bounce, where it acquires a non-singular value.}
\end{figure}

Figures \ref{fig1} and \ref{fig2} illustrate the bouncing behavior of, respectively, the scale factor and the Ricci scalar $R=6(\ddot{a}/a+\dot{a}^2/a^2)$ for radiation and for a large-$p$ example. In the first case, we used the special form of Kummer's function $_1F_1(-1/2;\,1/2;\,-z^2)=\rme^{-z^2}+\sqrt{\pi}\,z\,{\rm erf}(z)$, where erf is the error function:
\be
a(t) = \sqrt{\frac{2 \rme^{-\frac{1}{4} \Lambda ^2 t^2}}{\sqrt{\pi} \,\Lambda t_{\rm i}} +  
\frac{t}{t_{\rm i}} \, \text{erf} \left(\frac{\Lambda  t}{2}\right)}\,.
\ee

\begin{figure}
\hspace{-0.4cm}
\centering
\includegraphics[width=7.53cm]{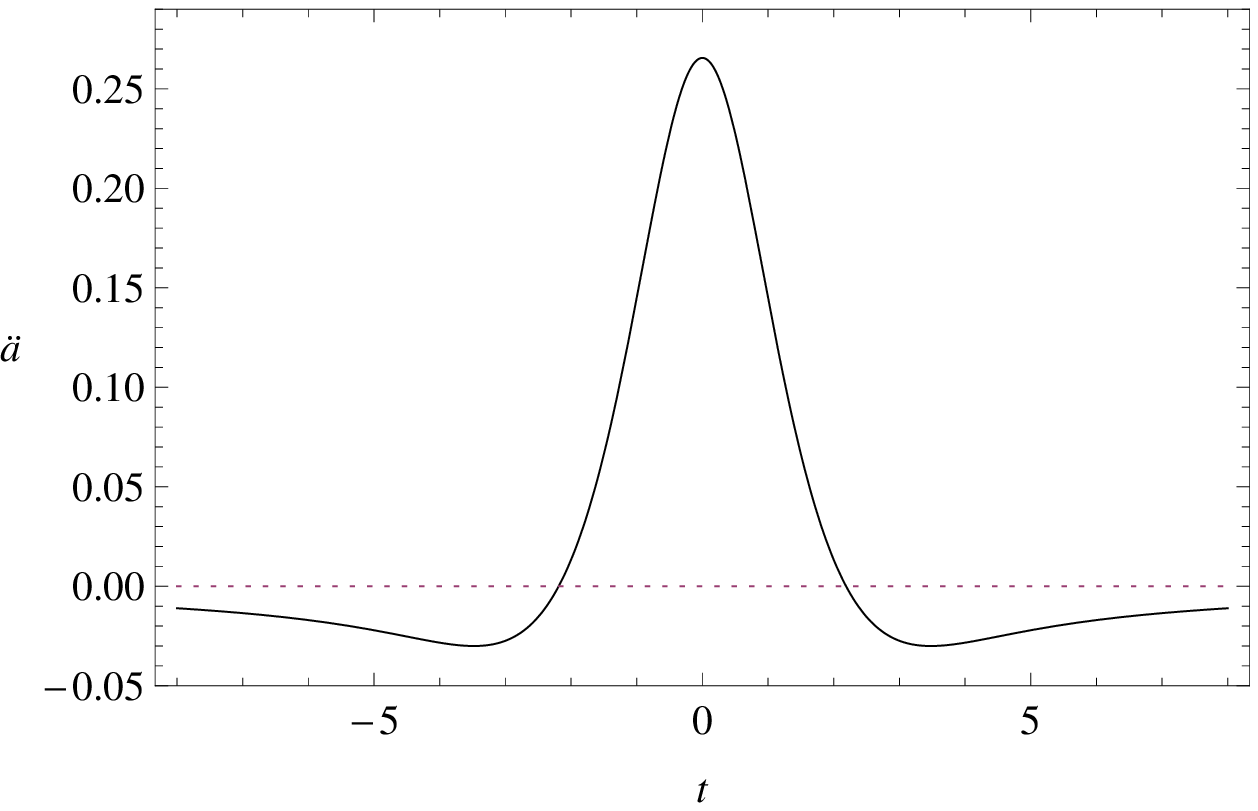}
\hspace{0.1cm}
\includegraphics[width=7.18cm]{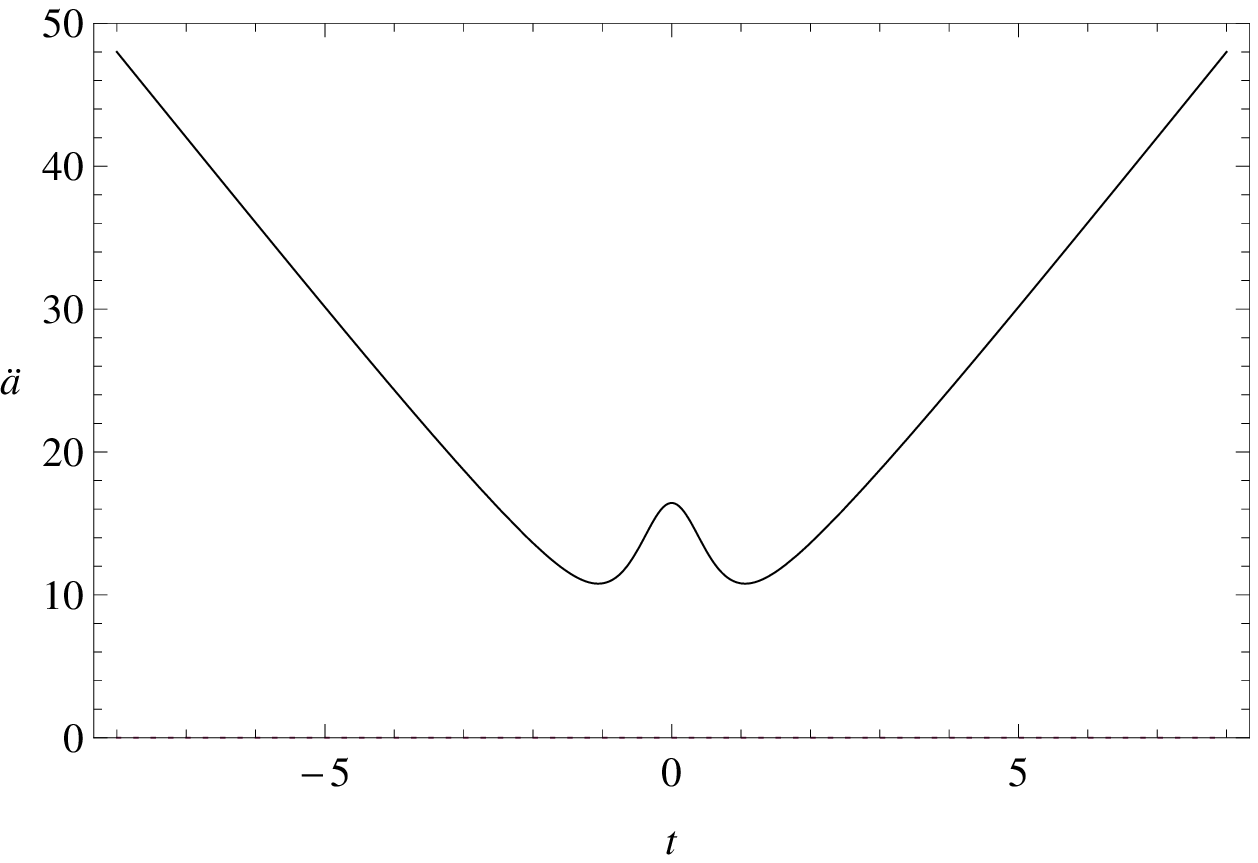}
\caption{\label{fig3} $\ddot{a}(t)$ for $p=1/2$ (top) and $p=3$ (bottom). In the presence of non-accelerating matter content (in this case, late-time radiation), when $\ddot a=0$ the (super-)inflationary era near the bounce naturally ends.}
\end{figure}

A consequence of Eq.\ \Eq{solu} is that one always has super-acceleration near the bounce independently of the value of $p$. This mechanism of super-inflation, which does not need any slow-rolling or exotic (e.g., phantom) scalar field, may be viewed as due to the vacuum energy associated with the graviton fluctuation, which is present for any type of matter content. In fact, in our model there is no ghost involved in the regularization of the solution. The singularity removal is due only to the form of the graviton propagator and not to other physical or unphysical degrees of freedom. While any local higher-derivative theory of gravity typically shows ghost-like degrees of freedom, in the present case the non-local nature of the propagator gives a regular bouncing solution without introducing ghosts. Figure \ref{fig3} shows positivity of the acceleration $\ddot a$, and a graceful exit from inflation when the matter content obeys the classical dominant energy condition (i.e., non-inflationary matter).

The Hubble parameter associated with the solution \Eq{solu} is
\be\label{H2}
H(t)=\frac{\dot a}{a}=\frac{p \Lambda^2 t}{2} \frac{ _1F_1\left(1-p;\frac{3}{2};-\frac{1}{4} t^2 \Lambda ^2\right)}{_1F_1\left(-p;\frac{1}{2};-\frac{1}{4} t^2 \Lambda ^2\right)}\,.
\ee
At the bounce, $H=0$. Asymptotically,
\be\label{bounH}
H(t)\ \stackrel{\text{\tiny $t\to\pm\infty$}}{\sim}\ \frac{p}{t}\,, \qquad H(t)\ \stackrel{\text{\tiny $t\to 0$}}{\sim}\ \frac{p \Lambda^2}{2}t\,,
\ee
and, integrating at small times, we have
\be
\label{superacc}
a(t)\sim  \exp\left(\frac{p}{4}\Lambda^2 t^2\right) \qquad \mathrm{as} \qquad\ t\to 0\,,
\ee
thus getting the asymptotic behavior of the bounce in a neighborhood of $t=0$. The scale factor $a(t)$ for small times has a typical super-acceleration profile. This reminds one of an early contribution on cosmology in higher-derivative gravity \cite{Pol88}, even if, as just stated, in our case we do not need to invoke the inflaton.

We conclude this section with three comments. First, we have not checked the stability of the profile \Eq{solu}. In the case of general non-polynomial theories, it is not obvious whether non-locality may trigger potentially dangerous instabilities. This is not so in the present case, where the non-local operator is the exponential of the d'Alembertian. There is reiterated evidence in the literature that such a strongly damping form factor actually improves any stability-related problem (exponential non-locality is under much greater control than other non-local models; see, e.g., \cite{BK1,cuta3}). Moreover, stability of the solution can be checked by looking at the behavior at early and late times separately (the complete perturbation would then be given by joining the two asymptotic behaviors). At late times, however, our solution reduces to standard power-law cosmology with standard dynamical equations, whose properties are well known. Therefore, one would need to verify stability only in the asymptotic-freedom regime. In this limit, however, the dynamical equations are linear in $a^2$ and the perturbation analysis becomes trivial. Therefore, the linear homogeneous perturbation $\de(a^2)\approx 2 a(t)$ $\times\de a(t)$ of the solution $a(t)$ obeys the same linearized equation of the background (with unperturbed differential operators), $\de a\propto a$, and the perturbed background $a+\de a\propto a$ is only a physically irrelevant re-normalization of the scale factor. As a side remark, there should be no instability issues generated by non-locality even at the level of inhomogeneities. This is expected on the grounds that non-local theories with entire functions, such as exponential operators, do not introduce ghost, tachyon or Laplacian instabilities. We do not expect these arguments to be altered by a full calculation, which could be performed only with the knowledge of the full dynamics. This goes beyond the goals of the present paper.

\begin{figure}
\centering
\includegraphics[width=8.5cm]{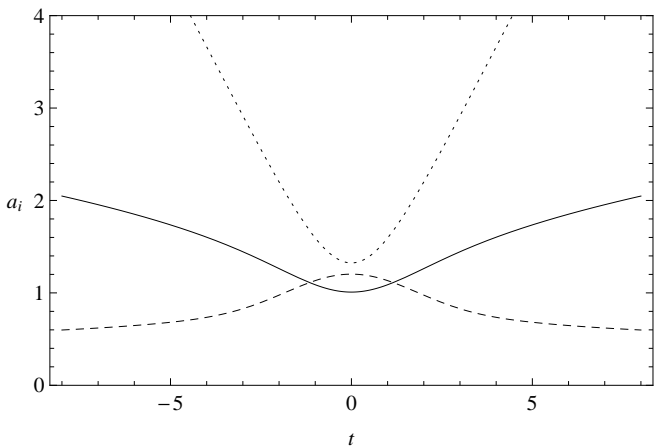}
\caption{\label{fig4} The three scale factors $a_i(t)$ of a Kasner solution with $p_1=-1/4$ (dashed curve), $p_2=(5-\sqrt{5})/8\approx 0.35$ (solid curve), and $p_3=(5+\sqrt{5})/8\approx 0.90$ (dotted curve), with $\Lambda=1=t_{\rm i}$.}  
\end{figure}

Second, Kasner solutions are also straightforward. In ordinary Einstein--Hilbert gravity, the solution for a flat homogeneous anisotropic universe is $a_i(t)=|t/t_{\rm i}|^{p_i}$, where $i=1,2,3$ and the exponents $p_i$ obey the two conditions $\sum_i p_i=1=\sum_i p_i^2$. Following the above procedure (see \cite{Bro11} for Kasner solutions in a local quantum-gravity model), from the propagator we get three copies of Eq.\ \Eq{solu}. This formula is automatically well-defined for all the values of $p_i$ allowed by the Kasner conditions. Each scale factor reaches its local extremum (a minimum for two $p_i>0$ and a maximum for $p_i<0$) at different values dictated by Eq.\ \Eq{boun}. Going forwards in time towards $t=0$ and beyond, two directions contract, undergo a bounce, and begin to expand. In the meanwhile, the third direction expands from a minimum value $a_{i*}$, reaches a maximal extension at the bounce, and then contracts back to $a_{i*}$. Figure \ref{fig4} shows one such solution.

Finally, we also checked that, in models where a Wick rotation is not necessary, the bounce picture persists. Let us recall that a Wick rotation was required because, in momentum space, the form factor \Eq{VSFT} is not convergent when integrating in $k^0$, $V(k^2)=\exp(\Lambda^2k^2)=\exp[\Lambda^2(k_0^2-|\vec{k}|^2)]$. On the other hand, using even powers of the Laplace--Beltrami operator renders the form factor convergent without transforming to imaginary time, and the bouncing-accelerating scenario still holds. For instance, in Krasnikov's model with the operator $V(\B)=\rme^{-\B^2/\Lambda^4}$, the solution with power-law asymptotic limit is a superposition of generalized hypergeometric functions $ _qF_s$, bounded from below (respectively, above) for $p>0$ ($<0$) by a symmetric bounce at some $a_*\neq 0$.


\subsection{Alternative derivation of the solution}\label{altder}

The solution \Eq{solu} can also be found via the diffusion-equation method \cite{cuta7}, which has proven to be a powerful tool both to address the Cauchy problem in exponen-tial-type non-local systems \cite{cuta3} and to find non-pertur-bative tachyon solutions in string theory at the level of target actions \cite{cuta4,cuta7,roll,cuta5}. While this method works well for matter fields on Minkowski background, the diffusion equation becomes non-linear when applied to the metric itself, since its solution appears also in the Laplace--Beltrami operator. For this reason, a particular gravitational action was constructed which allowed one to circumvent this problem \cite{cuta8}. Here, on the other hand, the action \Eq{Seff} gives rise to non-local, non-linear equations of motions for which the diffusion approach seems unsuitable. Fortunately, asymptotic freedom guarantees a window of applicability of the method in the limit where all interactions, including those of gravity with itself, are negligible.

Linearizing the action up to second order in perturbative fluctuations, one ends up with Eq.\ \Eq{kinnico}. Here, however, we do not invoke perturbation theory and throw away self- and matter interactions, but we encode them into an effective mass $m_{\rm eff}$ for the gravitational field. The latter obeys the equation of motion $\Box \, \rme^{\Box/\Lambda^2}\, h_{\mu \nu}+m_{\rm eff}^2h_{\mu \nu}=0$. To solve this equation (or, approximately, its non-linear extensions), we promote $h_{\mu\nu}(x)\to h_{\mu\nu}(x,r)$ to a field living in $D+1$ dimensions, where $r$ is an artificial extra direction (dimensionally, a squared length), and assume that $h$ obeys the diffusion equation
\be
(\p_r-\B)h_{\mu\nu}(x,r)=0\,,\qquad h_{\mu\nu}(x,0)=h_{\mu\nu}^{\rm cl}(x)\,,
\ee
with a given set of initial conditions $h_{\mu\nu}^{\rm cl}(x)$. Here $\B$ is the Laplace--Beltrami operator in the background metric, i.e., $\B=\eta^{\mu\nu}\p_\mu\p_\nu$ in our case. Once the solution is found, the parameter $r$ is fixed at some constant value $r=r_*$ such that the equation of motion $\Box h_{\mu \nu}(x,r_*+1/\Lambda^2)+m_{\rm eff}^2h_{\mu \nu}(x,r_*)=0$ is solved. 

For a homogeneous setting, the problem is drastically simplified. The only non-vanishing components of the metric are the diagonal spatial ones, and we need only to consider one diffusion equation for $a^2(t,r)$, with Laplace--Beltrami operator $\B=\p_t^2$ (spatial derivatives are immaterial). The initial condition at $r=0$ is nothing but the asymptotic classical profile at $t\to\pm\infty$ because the solution only depends on the ratio $-t^2/(4r)$ (this can be checked either {a posteriori} or beforehand by a simple scaling argument). Given a power-law initial condition, the solution of the diffusion equation
\be\label{die}
(\p_r-\p_t^2)a^2(t,r)=0\,,\qquad a^2(t,0)=a^2_{\rm cl}(t)=\left|\frac{t}{t_{\rm i}}\right|^{2p},
\ee
is, when evaluated at $r=r_*$, Eq.\ \Eq{solu} with $\Lambda^2=1/r_*$. The proof of this statement is essentially the calculation in \cite{cuta2} for a scalar-field profile in non-local cosmologies with exponential operators.\footnote{The reader can track down the steps in \cite{cuta3} from Eqs.\ (45) to (81), with the following mapping from the symbols used there to those adopted here: $\psi=\psi_3\to a^2$, $\phi_0\to t_{\rm i}^{-2p}$, $\theta/4\to p$, $\nu\to 1/2$, $H_0\to 0$, $\a\to -1$. In \cite{cuta3}, the opposite convention $\eta={\rm diag}(-,+,\cdots,+)$ is used for the spacetime signature.} Summarizing the procedure in a nutshell, one expands the initial condition $a^2(t,0)$ as an integral superposition of the eigenfunctions $\rme^{\pm\rmi E t}$ of the exponential operator $V^{-1}(\B)$. Applying $V^{-1}(\B)$ to the initial condition and performing the integration, one obtains a linear combination $C_1 a^2_1(t,r)+C_2a^2_2(t,r)$ of the two solutions to the second-order equation \Eq{die},  where
\ba
a_1^2(t,r)&=& \frac{2\Gamma(1+p)}{\sqrt{\pi}}\,
\left(\frac{4r}{t_{\rm i}^2}\right)^p \sqrt{\frac{t^2}{4r}}\nonumber\\
&&\times {}_1F_1\left(\frac12-p;\frac32;\,-\frac{t^2}{4r}\right)\,,\label{got}\\
a_2^2(t,r)&=& \left(-\frac{4r}{t_{\rm i}^2}\right)^p \Psi\left(-p;\frac12;\,-\frac{t^2}{4r}\right)\,,
\ea
and $\Psi$ is the confluent hypergeometric function of the second kind. The solution $a_1^2$ has a big-bang singularity and is not well-defined (positive definite, finite, and so on) for all values of $p>0$; a special case is $p=1/2$, for which $a_1(t)=\sqrt{|t/t_{\rm i}|}$. The solution $a_2^2$ is complex-valued for general $p$; for $p=1$, however, we get $a_2(t)= \sqrt{2r/t_{\rm i}^2+(t/t_{\rm i})^2}$, which is real-valued and bouncing for $r\neq 0$, and singular if one takes trivial diffusion ($r=0$, also found in \cite{Kos13}). The choice of coefficients $C_1 = -\rmi \tan(\pi p)$ and $C_2 = \rme^{-\rmi\pi p}/\cos(\pi p)$ gives (the square of) our solution \Eq{solu}. This linear combination is real-valued, big-bang free, super-accelerating near the bounce, and valid for all positive $p$. It is the latter criterion which makes us believe that bouncing solutions are typical in this theory, despite the existence of non-bouncing very special cases such as those above.


\subsection{Effective dynamics}

\begin{figure}
\centering
\includegraphics[width=7.53cm]{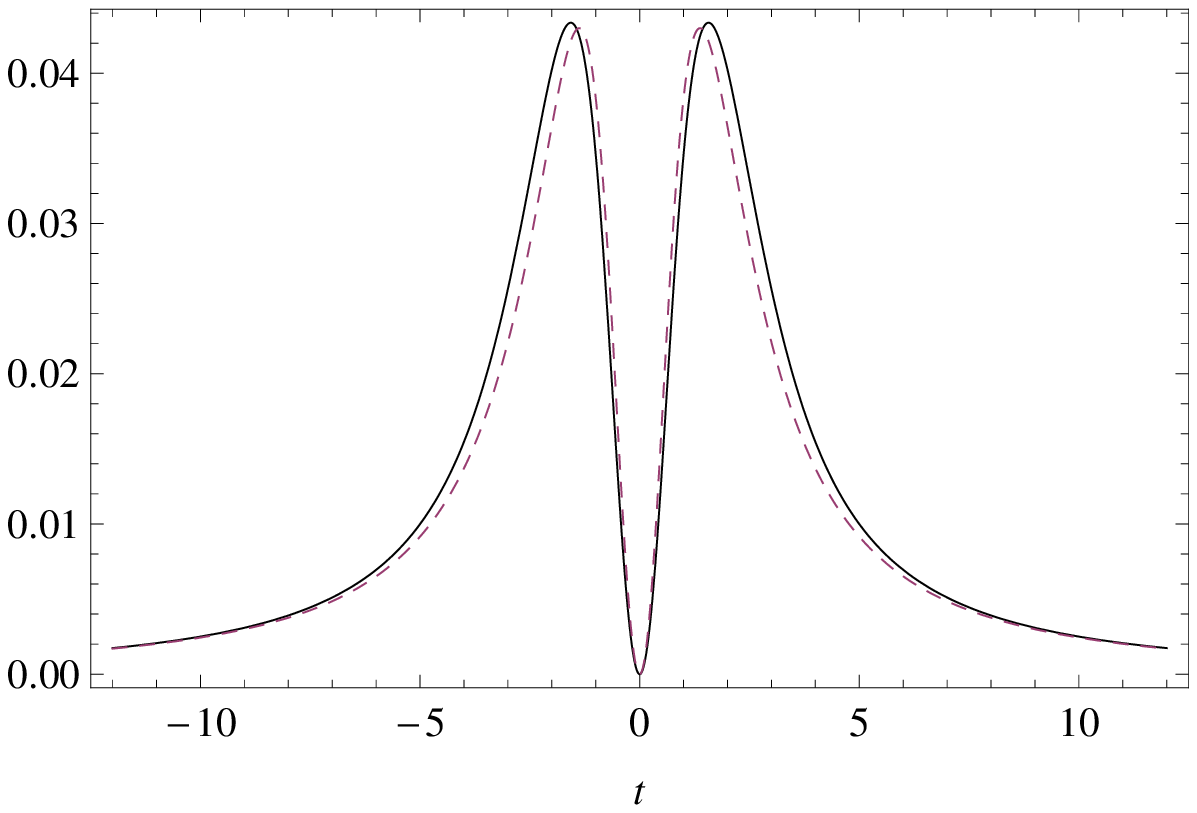}
\includegraphics[width=7.4cm]{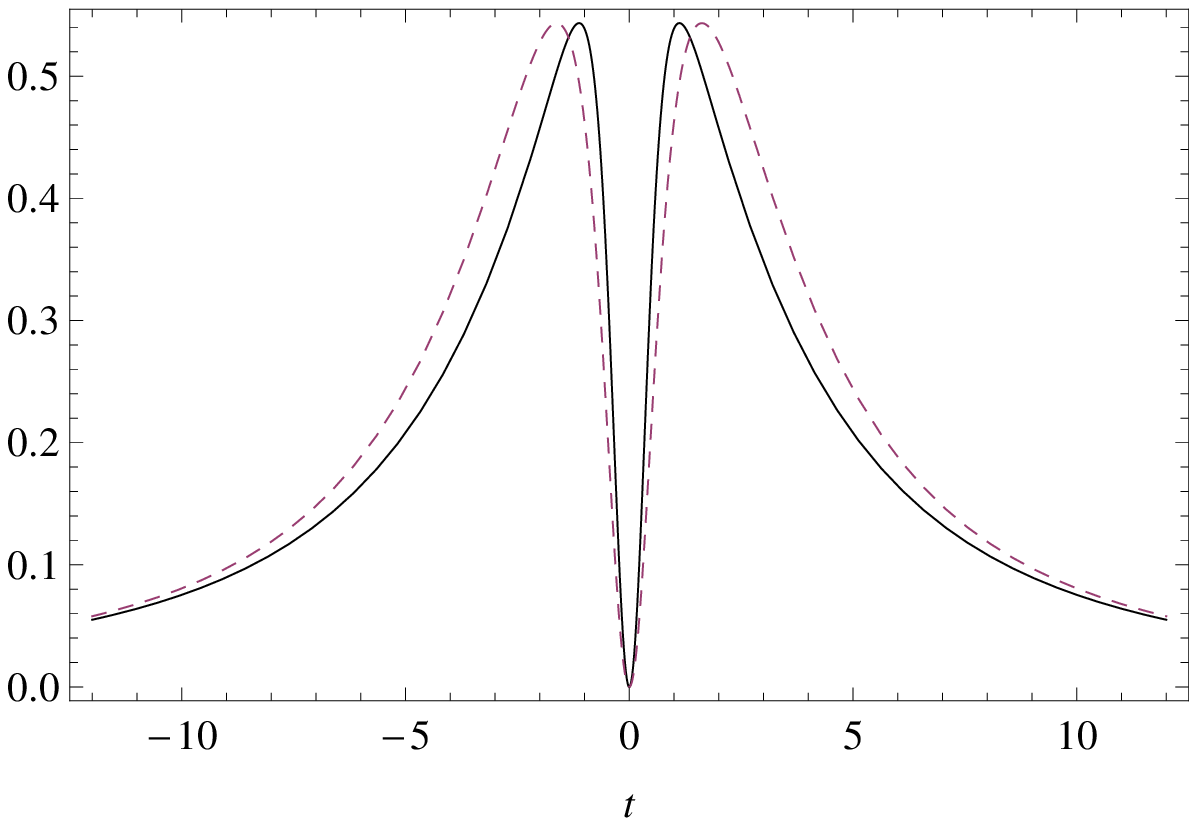}
\caption{\label{fig5} Plot of $H^2$ (Eq.\ \Eq{H2}, solid curve) and the effective energy density \Eq{reff} (dashed line), for $p=1/2$, $\b= 0.83$ (upper plot) and $p=3$, $\b= 1.3$ (lower plot), with $\Lambda=1$.}
\end{figure}

The solution \Eq{solu} is approximate. Thanks to asymptotic freedom, it is valid at early times $t\lesssim t_{\rm i}$ where interactions are negligible. It is valid also at late times, where the theory reduces to ordinary Einstein--Hilbert gravity plus sub-leading quadratic terms (Stelle model). To check its viability outside these regimes, it is interesting to fit the Hubble parameter \Eq{H2} with the effective energy density $\rho_{\rm eff}$ defining the modified Friedmann equation \Eq{freq}. In the right-hand side of that equation, we take the energy density $\rho$ to be the one of a standard power-law cosmology, but with $a=|t/t_{\rm i}|^p$ replaced by Eq.\ \Eq{solu},
\be\label{ro}
\rho(t) = \frac{3p^2}{\k^2 t_{\rm i}^2}\frac{1}{a^{2/p}(t)}\,,
\ee
while the critical energy density  $\rho_*\neq \rho_{\rm eff}(0)$ is fixed by the bounce scale factor $a_*=a(0)$ in \Eq{boun},
\ba
\rho_* &=& \frac{3p^2}{\k^2 t_{\rm i}^2}\frac{1}{a_*^{2/p}}= \frac{3p^2\Lambda^2}{4\k^2}\left[\frac{\sqrt{\pi}}{\Gamma\left(\frac12+p\right)}\right]^{\frac1p}\nonumber\\
&=&\frac{3p^2}{32\pi}\left[\frac{\sqrt{\pi}}{\Gamma\left(\frac12+p\right)}\right]^{\frac1p}\frac{\Lambda^2}{m_{\text{\tiny Pl}}^2}\rho_{\text{\tiny Pl}}\,,\label{rpl}
\ea
where $m_{\text{\tiny Pl}}=\sqrt{\hbar c/G}\approx 1.2209\,\times\,10^{19}\,\mbox{GeV}$ is the Planck mass and $\rho_{\text{\tiny Pl}}=m_{\text{\tiny Pl}}^4\approx 2.2\,\times\,10^{76}\,\mbox{GeV}^4$ is the Planck energy density. Overall,
\be\label{reff}
\frac{\k^2}{3}\rho_{\rm eff}(t)= \frac{p^2}{t_{\rm i}^2 a^{2/p}(t)}\left\{1-\left[\frac{a_*}{a(t)}\right]^{\frac{2\b}{p}}\right\}\,,
\ee
which is the quantity plotted in Fig.\ \ref{fig5}. For $p=1/2$ (radiation case), we achieved a good qualitative fit with $\b\approx 0.83$, while for $p=3$ we plot $\b\approx 1.3$.

Notice that there may be no compelling reason to impose an all-scale fitting as in the figure. Using the asymptotic limits \Eq{largez}, we have Eq.\ \Eq{bounH} and
\ba
&&\sqrt{\frac{\k^2}{3}\rho_{\rm eff}(t)}\ \stackrel{\text{\tiny $t\to\pm\infty$}}{\sim}\ \frac{p}{t}\,,\nonumber\\
&&\sqrt{\frac{\k^2}{3}\rho_{\rm eff}(t)}\ \stackrel{\text{\tiny $t\to 0$}}{\sim}\ \sqrt{\frac{2\b}{t_{\rm i}^2a_*^{2/p}\Lambda^2}}\frac{p \Lambda^2}{2}t\,.\label{bounr}
\ea
The relative error at the origin is minimized for 
\be\label{beta}
\b=a_*^{\frac2p}\frac{\Lambda^2t_{\rm i}^2}{2}=2\left[\frac{\Gamma\left(\frac12+p\right)}{\sqrt{\pi}}\right]^{\frac1p},
\ee
equal to $\b\approx 0.64$ for $p=1/2$ and $\b\approx 2.47$ for $p=3$. For these values (which neither correspond to the fit of the figure nor to a least maximal error in the fit), there is a significant difference in the height of the local symmetric maxima at intermediate times. Conversely, in the plots the maximal relative error occurs at the origin. We believe that Eq.\ \Eq{beta} better represents the asymptotic solutions, since higher-order curvature and quantum corrections are expected anyway to modify the evolution at mesoscopic scales, where the two lumps are located.

\begin{figure}
\hspace{-0.6cm}
\centering
\includegraphics[width=7.4cm]{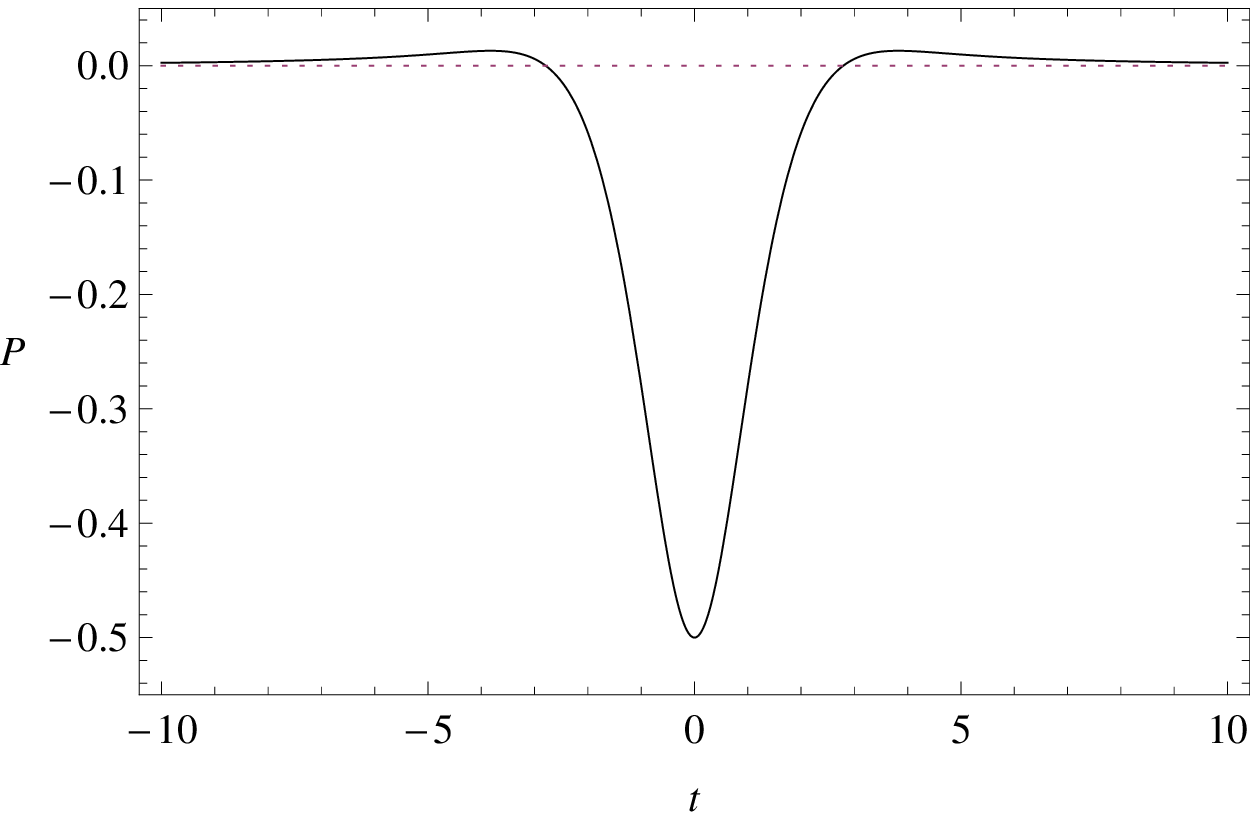}
\hspace{0.2cm}
\includegraphics[width=7.4cm]{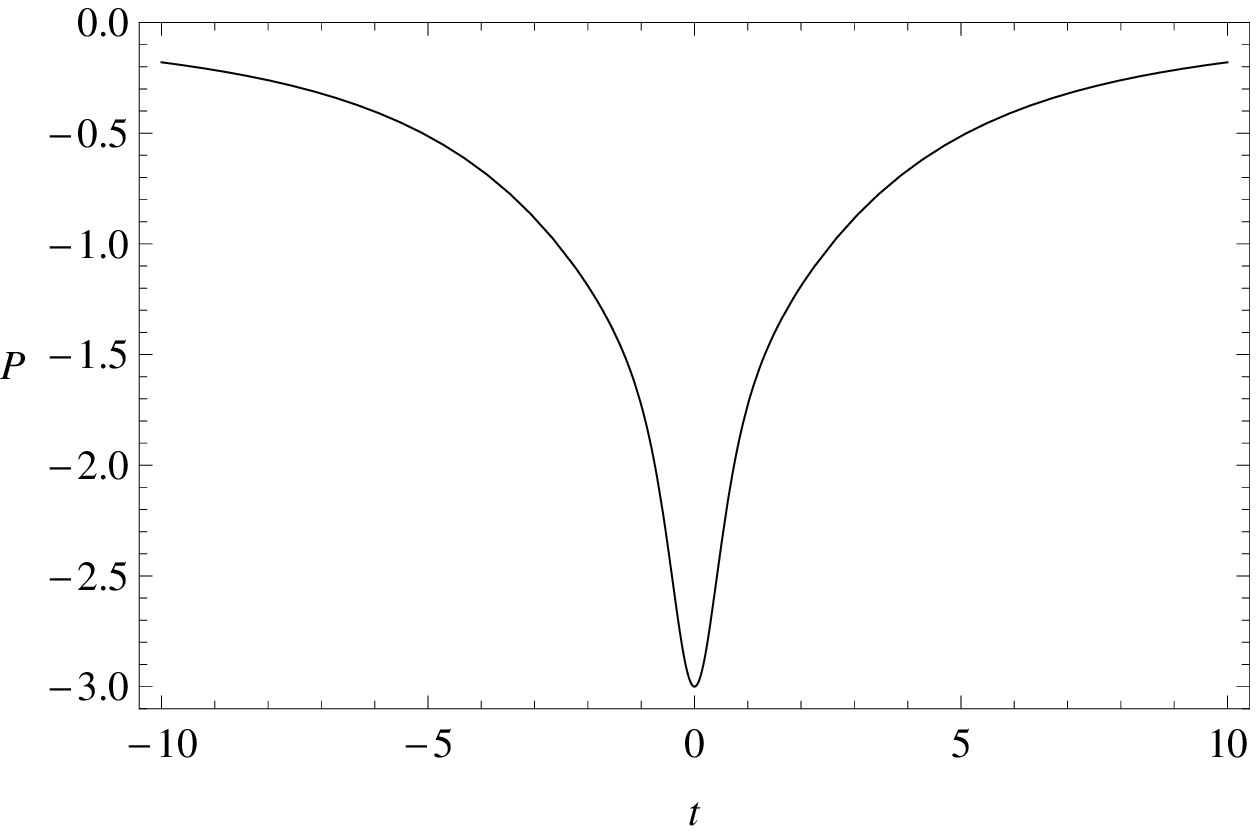}
\caption{\label{fig6} The effective pressure $P(t)=-\rho_{\rm eff}-\dot\rho_{\rm eff}/(3H)$ for $p=1/2$ (top) and $p=3$ (bottom). At the bounce, it tends to a non-zero finite value $P(0)=-R(0)/(3\k^2)$. Here we set $\k^2=1=\Lambda$.}
\end{figure}

\begin{figure}
\hspace{-0.6cm}
\centering
\includegraphics[width=7.2cm]{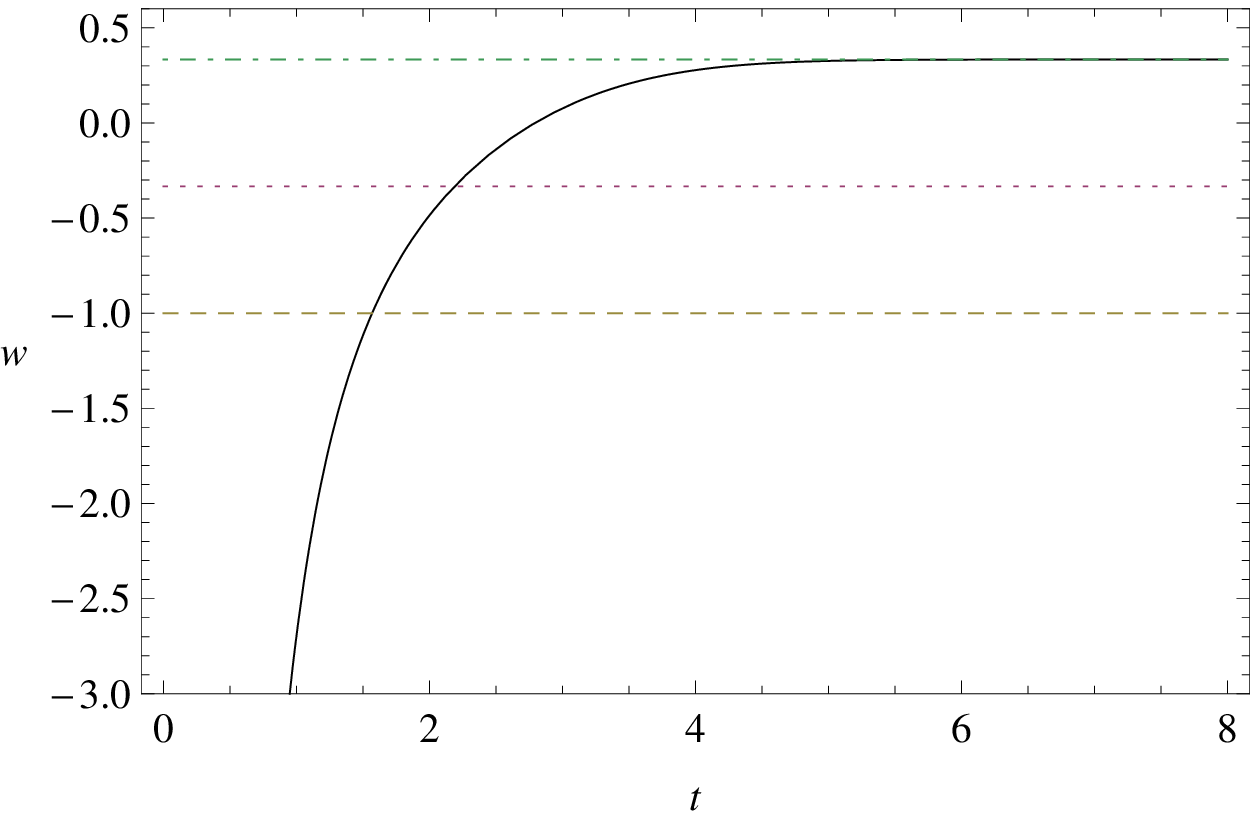}
\hspace{0.2cm}
\includegraphics[width=7.3cm]{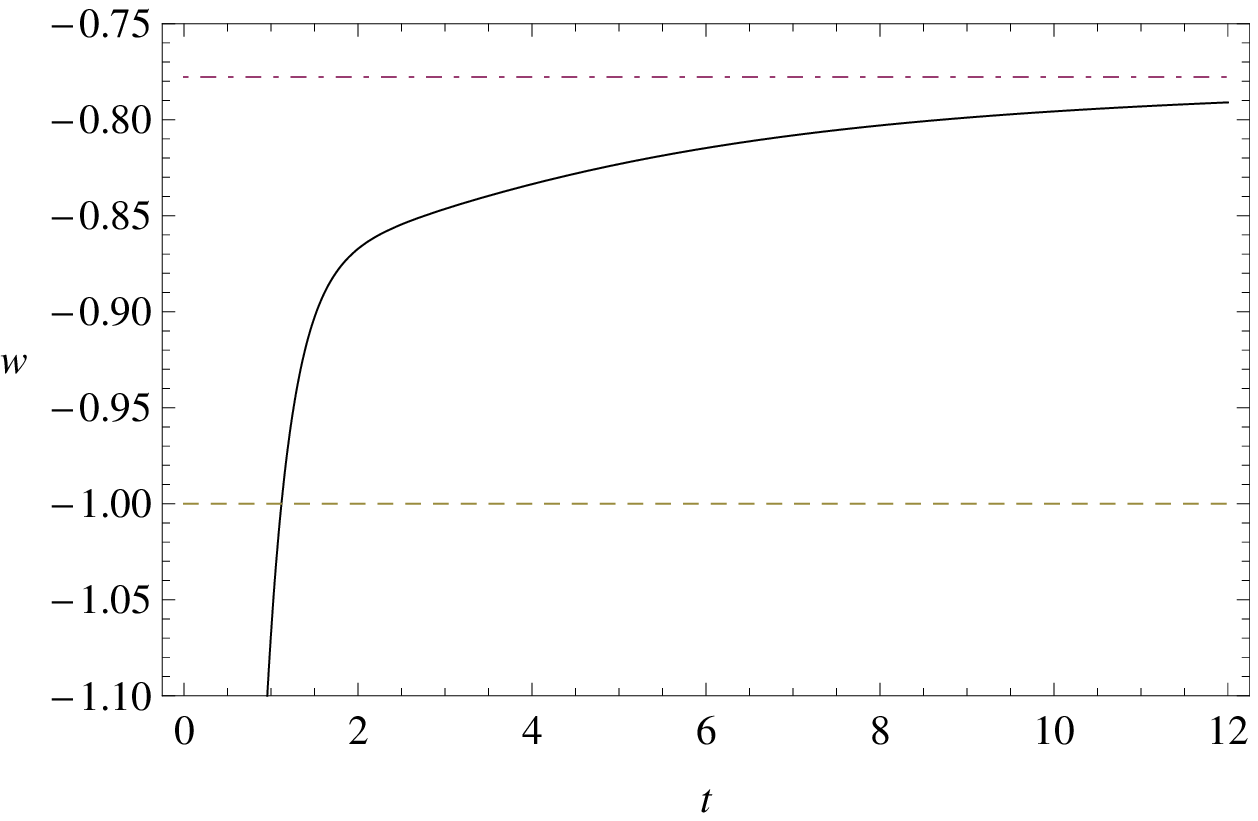}
\caption{\label{fig7} The effective barotropic index $w(t)=-P(t)/\rho_{\rm eff}(t)$ for $p=1/2$ (top) and $p=3$ (bottom), with $\Lambda=1$. For $w<-1/3$, one has acceleration and for $w<-1$ (small times) super-acceleration, in accordance with Eq.\ (\ref{superacc}). The lines $w=-1+2/(3p)$ (dot-dashed), $w=-1/3$ (dotted) and $w=-1$ (dashed) are shown for reference.}
\end{figure}

Assuming a standard Raychaudhuri equation $\dot\rho_{\rm eff}+3H(P+\rho_{\rm eff})=0$, one can also obtain an effective pressure $P$ and barotropic index $w$,
\ba
&& P(t):=-\rho_{\rm eff} -\frac{\dot{\rho}_{\rm eff}}{3H}=-\frac{1}{\k^2}\left(H^2+2\frac{\ddot a}{a}\right)\,,\nonumber\\
&& w(t):=\frac{P}{\rho_{\rm eff}}\,,\label{Peff}
\ea
which are depicted in Figs.\ \ref{fig6} and \ref{fig7}, respectively. At the bounce, one has a non-vanishing finite effective pressure $P(0)=-R(0)/(3\k^2)$, where $R(0)=3p\Lambda^2$ is the Ricci scalar. At late times, $w\to-1+2/(3p)$, while at early times
\be
w(t)\ \stackrel{t\to 0}{\sim}\ -1-\frac{4}{3p\Lambda^2}\frac{1}{t^2}\,,
\ee
where one has super-acceleration ($w<-1$).


\section{Conclusions}

In this paper, we have proposed a non-local model of gravity with improved UV behavior. We have focused only on its cosmology, and used the property of asymptotic freedom to find approximate solutions valid both at early and late times. In general, these solutions possess a bounce and avoid the big-bang singularity, and have an early era of acceleration with a natural exit in the absence of inflaton fields. Specifically, the universe is characterized by a super-acceleration regime at the bounce, with effective barotropic index $w<-1$. 

Since we have not solved the full equations of motion exactly, we have bypassed the problem of getting the dynamics of a theory with all sectors (both gravity and matter) being non-local. Instead, we have matched the solution \Eq{solu} with an effective dynamics encoded in the modified Friedmann equation \Eq{freq}. The good agreement between this solution and Eq.\ \Eq{freq} at all scales suggests that the problem is not unsolvable. The diffusion-equation method, applied in Sect.\ \ref{altder} to find an alternative derivation of the profile \Eq{solu}, might be a useful tool in this respect. Intriguingly, Eq.\ \Eq{die} acts as a ``beta function,'' determining the running of the metric with the cut-off length scale $1/\Lambda$, which plays the role of diffusion time $\sim \sqrt{r}$. The possibility to study the dynamics of this class of non-local theories via the diffusion method is a direct consequence of their renormalization properties \cite{cuta7}.

It is remarkable that the bouncing dynamics of the present model can be reproduced semi-quantitatively by an effective equation with only one free parameter. The value of the critical energy density $\rho_*$ is also plausible. From Eq.\ \Eq{rpl} and setting the cut-off to its natural value $\Lambda=m_{\text{\tiny Pl}}$, we get $\rho_*\approx 0.02\,\rho_{\text{\tiny Pl}}$ for $p=1/2$ and $\rho_*\approx 0.22\,\rho_{\text{\tiny Pl}}$ for $p=3$. Models such that $\rho\leq\rho_*\leq \rho_{\text{\tiny Pl}}$ must have $p\leq p_{\rm max}\approx 12.67$. For $p> p_{\rm max}$, the critical energy density exceeds $\rho_{\text{\tiny Pl}}$, and the energy density of the universe can become trans-Planckian near and at the bounce. Therefore, scenarios with too-large $p$ are not well represented by Eq.\ \Eq{freq}. This is not an issue in our model, since we have early-universe acceleration by default also for small values of $p$.

The relative error between the approximate solution $H^2$ and the effective energy density $\k^2\rho_{\rm eff}/3$ could be reduced by a more refined \emph{Ansatz} for the effective Friedmann equation. Further study of this method may turn out to shed some light on the exact dynamics, which should be developed in parallel starting from the actual equations of motion. For instance, once derived the actual equations of motion of the theory one could plug the solution $a(t)$ found here, and check whether a reasonable matter sector is recovered. In particular, from our modified Kasner solution it should be possible to check whether some form of BKL chaos survives in the full anisotropic dynamics.

This should not only clarify whether the bounce picture is robust in our theory, but also to which class of singularity-free cosmologies the model belongs to. As an exact dynamical equation, expression \Eq{freq} with $\b=1$ appears also in braneworlds with a timelike extra direction \cite{ShS} and in purely homogeneous loop quantum cosmology (for reviews consult, e.g., \cite{lqcr1,lqcr2}) only in the parameter choice for the so-called ``improved'' mini-superspace dynamics \cite{sin06,APS} (for other parametrizations, Eq.\ \Eq{freq} with $\b=1$ no longer holds \cite{CH}). Although there is no relation between our framework and these high-energy cosmological models, they all share the same type of bounce where the right-hand side of the Friedmann equation receives a negative higher-order correction in the energy density. On the other hand, there is a different class of models where the correction is of the ``dark radiation'' form $-1/a^4$, which is responsible for the bounce at $H=0$. Such is the case for the Randall--Sundrum braneworld with a spacelike extra dimension \cite{BDEL}, Ho\v{r}ava--Lifshitz gravity without detailed balance \cite{lif1,KiK}, and cosmologies with fer-mionic condensates \cite{AB3,ABC}. As far as we pushed the analysis in this paper, the present model apparently lies in the first category, with the added bonus that we have an alternative mechanism of inflation of purely geometric origin.


\begin{acknowledgements}
The authors thank G.\ Mena Marug\'an for comments on the manuscript and are grateful to J.\ Moffat for early discussions on related topics. G.C.\ and L.M.\ acknowledge the i-Link cooperation programme of CSIC (project ID i-Link0484) for partial sponsorship. The work of G.C.\ is under a Ram\'on y Cajal tenure-track contract; he also thanks Fudan University for the kind hospitality during the completion of this article. The work of P.N.\ has been supported by the German Research Foundation (DFG) grant NI 1282/2-1, and partially by the Helmholtz International Center for FAIR within the framework of the LOEWE program (Landesoffensive zur Entwicklung Wissenschaftlich-\"{O}konomischer Exzellenz) launched by the State of Hesse and by the European COST action MP0905 ``Black Holes in a Violent Universe''.
\end{acknowledgements}


\end{document}